\documentclass[runningheads]{llncs}
\usepackage{graphicx}
\usepackage{amsmath,amssymb,amsfonts,bm}
\usepackage[export]{adjustbox} 
\usepackage[ruled,vlined]{algorithm2e}
\usepackage{hyperref}
\usepackage[caption=false]{subfig}
\usepackage{color,soul}
\usepackage[table,xcdraw]{xcolor}
\usepackage{multirow}
\usepackage{cite}
\usepackage{siunitx}
\usepackage[utf8]{inputenc}
\usepackage[strings]{underscore} 
\usepackage[english]{babel} 
\usepackage{duckuments}
\usepackage[inline]{enumitem}
\newlist{inlinelist}{enumerate*}{1} 
\setlist[inlinelist]{label=(\roman*)}

\setlist[itemize]{nosep} 

\usepackage{pifont}
\newcommand{\cmark}{\text{\ding{51}}} 
\newcommand{\xmark}{\text{\ding{55}}} 

\usepackage[]{threeparttable}

\usepackage{floatrow}
\usepackage{wrapfig}

\begin{document}
%
\title{Reducing Annotation Need in Self-Explanatory Models for Lung Nodule Diagnosis
}
%
\titlerunning{Reducing Annotation Need in Self-Explanatory Models}
\author{Jiahao Lu\inst{1,2}
\and Chong Yin\inst{1,3}
\and Oswin Krause\inst{1}
\and Kenny Erleben\inst{1}
\and Michael Bachmann Nielsen\inst{2}
\and Sune Darkner\inst{1}
}
\authorrunning{J. Lu et al.}



\institute{Department of Computer Science, University of Copenhagen, Denmark \and
Department of Diagnostic Radiology, Rigshospitalet, Copenhagen University Hospital, Denmark \and
Department of Computer Science, Hong Kong Baptist University, China}

\maketitle              
\begin{abstract}
Feature-based self-explanatory methods explain their classification in terms of human-understandable features. In the medical imaging community, this semantic matching of clinical knowledge adds significantly to the trustworthiness of the AI.
However, the cost of additional annotation of features remains a pressing issue. 
We address this problem by proposing cRedAnno, a data-/annotation-efficient self-explanatory approach for lung nodule diagnosis. 
cRedAnno considerably reduces the annotation need by introducing self-supervised contrastive learning to alleviate the burden of learning most parameters from annotation, replacing end-to-end training with two-stage training.
When training with hundreds of nodule samples and only $1\%$ of their annotations, cRedAnno achieves competitive accuracy in predicting malignancy, meanwhile significantly surpassing most previous works in predicting nodule attributes. 
Visualisation of the learned space further indicates that the correlation between the clustering of malignancy and nodule attributes coincides with clinical knowledge.
Our complete code 
is open-source available: \url{https://github.com/diku-dk/credanno}.

\keywords{Explainable AI \and Lung nodule diagnosis \and Self-explanatory model \and Intrinsic explanation \and Self-supervised learning.}
\end{abstract}

\section{Introduction}
\label{sec:intro}



\begin{figure}[tbp]
    \centering
    \subfloat[Previous methods]{
        \adjincludegraphics[height=0.5\textwidth,trim=0 0 {.65\width} 0,clip]{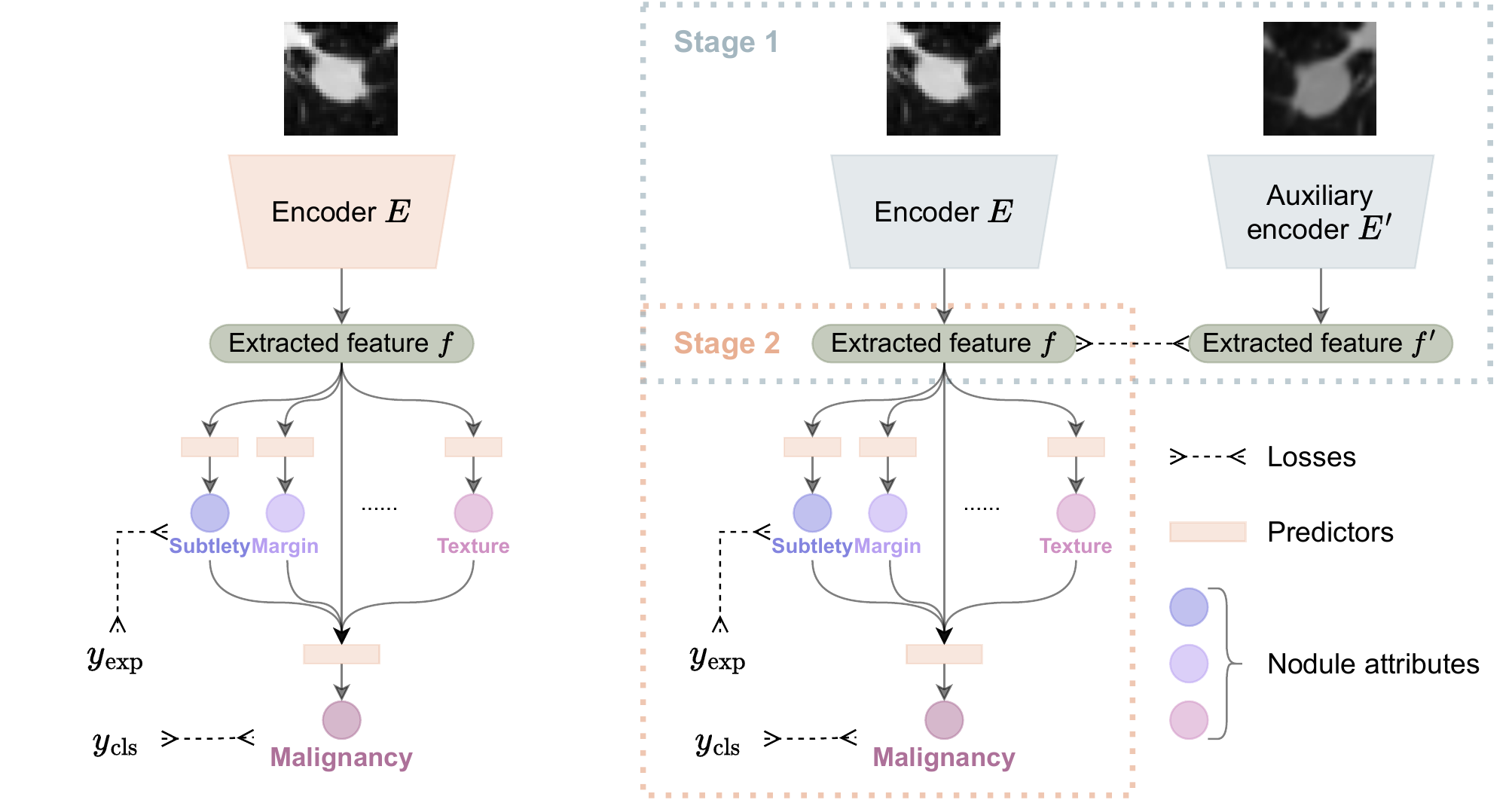}
        \label{fig:overview_a}}\hfil
    \subfloat[cRedAnno]{
        \adjincludegraphics[height=0.5\textwidth,trim={.38\width} 0 0 0,clip]{figs/intro.pdf}
        \label{fig:overview_b}}
    \caption{\textbf{Concept illustration.}
    (a) Previous works are trained end-to-end, where all parameters are learned from the annotations. 
    (b) Our proposed cRedAnno uses two-stage training, where most of the parameters are learned during the first stage in a self-supervised manner. Therefore, in the second stage, only few annotations are needed to train the predictors.
    }
    \label{fig:overview}
\end{figure}

Lung cancer is one of the leading causes of cancer deaths worldwide due to its high morbidity and low survival rate \cite{cielloMissedLungCancer2017}. 
In clinical practice, accurate characterisation of pulmonary nodules in CT images is an essential step for effective lung cancer screening \cite{vlahosLungCancerScreening2018}. 
Modern deep-learning-based ``black box" algorithms, although achieving accurate classification performance \cite{al-shabiLungNoduleClassification2019}, are hardly acceptable in high-stakes medical diagnosis\cite{vanderveldenExplainableArtificialIntelligence2022}.

Amongst recent efforts to develop explainable AI \cite{barredoarrietaExplainableArtificialIntelligence2020} to bridge this gap \cite{tjoaSurveyExplainableArtificial2021, vanderveldenExplainableArtificialIntelligence2022},
post-hoc approaches that attempt to explain such ``black boxes" are not deemed trustworthy enough \cite{rudinStopExplainingBlack2019}.
In contrast, feature-based self-explanatory methods are trained to first predict a set of well-known human-interpretable features, and then use these features for the final classification (Fig.~\ref{fig:overview_a})\cite{stammerRightRightConcept2021, shenInterpretableDeepHierarchical2019, lalondeEncodingVisualAttributes2020}.
This is believed to be especially valuable in medical applications because such semantic matching towards clinical knowledge tremendously increases the AI's trustworthiness \cite{salahuddinTransparencyDeepNeural2022}. 
Unfortunately, the required additional annotation on features still limits the applicability of this approach in the medical domain.

This paper aims to minimise additional annotation need for predicting malignancy and nodule attributes in lung CT images. 
We achieve this by separating the training of model's parameters into two stages, as shown in Fig.~\ref{fig:overview_b}. 
In Stage 1, the majority of parameters are trained using self-supervised contrastive learning \cite{heMomentumContrastUnsupervised2020, grillBootstrapYourOwn2020, caronEmergingPropertiesSelfSupervised2021} as an encoder to map the input images to a latent space that complies with radiologists' reasoning for nodule malignancy. 
In Stage 2, a small random portion of labelled samples is used to train a simple predictor for each nodule attribute. 
Then the predicted human-interpretable nodule attributes are used jointly with the extracted features to make the final classification.

Our experiments on the publicly available LIDC dataset \cite{armatoLungImageDatabase2011} show that 
with fewer nodule samples and only $1\%$ of their annotations, the proposed approach achieves comparable or better performance compared with state-of-the-art methods using full annotation \cite{shenInterpretableDeepHierarchical2019, lalondeEncodingVisualAttributes2020, chenEndtoEndMultiTaskLearning2021, liuMultiTaskDeepModel2020, joshiLungNoduleMalignancy2021}, and reaches approximately $90\%$ accuracy in predicting all nodule attributes simultaneously.
By visualising the learned space, the extracted features are shown to be highly separable and correlated well with clinical knowledge.

\section{Method}
\label{sec:method}

As the illustrated concept in Fig.~\ref{fig:overview_b}, the proposed approach consists of two parts: unsupervised training of the feature encoder and supervised training to predict malignancy with human-interpretable nodule attributes as explanations.

\subsubsection{Unsupervised feature extraction}

Due to the outstanding results exhibited by DINO \cite{caronEmergingPropertiesSelfSupervised2021}, we adopt their framework for unsupervised feature extraction, which trains 
\begin{inlinelist}
    \item a primary branch $\{E, H\}_{\theta_\text{pri}}$, composed by a feature encoder $E$ and a multi-layer perceptron (MLP) prediction head $H$, parameterised by $\theta_\text{pri}$;
    \item an auxiliary branch $\{E, H\}_{\theta_\text{aux}}$, which is of the same architecture as the primary branch, while parameterised by $\theta_\text{aux}$.
\end{inlinelist}
After training only the primary encoder $E_{\theta^\text{E}_\text{pri}}$ is used for feature extraction.

The branches are trained using augmented image patches of different scales to grasp the core feature of a sample.
For a given input image $x$, different augmented global views $V^g$ and local views $V^l$ are generated \cite{caronUnsupervisedLearningVisual2020}:
$x \rightarrow v \in V^g \cup V^l$.
The primary branch is only applied to the global views $v_\text{pri} \in V^g$, producing $K$ dimensional outputs $z_\text{pri}=E_{\theta^\text{E}_\text{pri}} \circ H_{\theta^\text{H}_\text{pri}} (v_\text{pri})$;  while the auxiliary branch is applied to all views $v_\text{aux} \in V^g \cup V^l$, producing outputs $z_\text{aux}=E_{\theta^\text{E}_\text{aux}} \circ H_{\theta^\text{H}_\text{aux}} (v_\text{aux})$ to predict $z_\text{pri}$.
To compute the loss, the output in each branch is passed through a Softmax function scaled by temperature $\tau_\text{pri}$ and $\tau_\text{aux}$:
$ p_\text{aux} = \texttt{softmax}(z_\text{aux} / \tau_\text{aux})$,\quad
$p_\text{pri} = \texttt{softmax}((z_\text{pri}-\mu) / \tau_\text{pri})$,
where a bias term $\mu$ is applied to $z_\text{pri}$ to avoid collapse\cite{caronEmergingPropertiesSelfSupervised2021}, and updated at the end of each iteration using the exponential moving average (EMA) of the mean value of a batch with batch size $N$ using momentum factor $\lambda \in [0, 1)$: $\mu \leftarrow \lambda \mu + (1 - \lambda) \frac{1}{N} \sum_{s=1}^{N} z_\text{pri}^{(s)}$.

The parameters $\theta_\text{aux}$ are learned by minimising the cross-entropy loss between the two branches via back-propagation \cite{heMomentumContrastUnsupervised2020}:
\begin{equation}
    \theta_\text{aux} \leftarrow 
    \arg\min_{\theta_\text{aux}} \sum_{v_\text{pri} \in V^g} \sum_{\substack{v_\text{aux} \in V^g \cup V^l \\ v_\text{aux} \neq v_\text{pri}}} \mathcal{L}\left( p_\text{pri}, p_\text{aux} \right) ,
\end{equation}
where $\mathcal{L}({p}_1, {p}_2) = - \sum_{c=1}^C {p}_1^{(c)} \log {p}_2^{(c)}$ for $C$ categories. 
The parameters $\theta_\text{pri}$ of the primary branch are updated by the EMA of the parameters $\theta_\text{aux}$ with momentum factor $m \in [0, 1)$:
\begin{equation}
    \theta_\text{pri} \leftarrow m\theta_\text{pri} + (1 - m)\theta_\text{aux} .
\end{equation}

In our implementation, the feature encoders $E$ use Vision Transformer (ViT)\cite{dosovitskiyImageWorth16x162020} as the backbone for their demonstrated ability to learn more generalisable features. Following the basic implementation in \texttt{DeiT-S}\cite{touvronTrainingDataefficientImage2021}, 
our ViTs consist of $12$ layers of standard Transformer encoders \cite{vaswaniAttentionAllYou2017} with $6$ attention heads each. 
The MLP heads $H$ consist of three linear layers (with GELU activation 
) with $2048$ hidden dimensions, followed by a bottleneck layer of $256$ dimensions, $l_2$ normalisation and a weight-normalised layer \cite{salimansWeightNormalizationSimple2016} 
to output predictions of $K = 65536$ dimensions, as suggested by \cite{caronEmergingPropertiesSelfSupervised2021}. 

\subsubsection{Supervised prediction}
After the training of feature encoders is completed, the learned parameters $\theta^\text{E}_\text{pri}$ in the primary encoder are frozen and all other components are discarded. Given an image $x$ with malignancy annotation $y_\text{cls}$ and explanation annotation $y_\text{exp}^{(i)}$ for each nodule attribute $i=1,\cdots, M$, its feature is extracted via the primary encoder: $f = E_{\theta^\text{E}_\text{pri}} (x)$.

The prediction of each nodule attribute $i$ is generated by a predictor $G_\text{exp}^{(i)}$:
$z_\text{exp}^{(i)} = G_\text{exp}^{(i)} (f)$.
Then the malignancy prediction $z_\text{cls}$ is generated by a predictor $G_\text{cls}$ from the concatenation ($\oplus$) of extracted features $f$ and predictions of nodule attributes:
\begin{equation}
    z_\text{cls} = G_\text{cls} (f \oplus z_\text{exp}^{(1)} \oplus \cdots \oplus z_\text{exp}^{(M)}).
\end{equation}
The predictors are trained by minimising the cross-entropy loss between the predictions and annotations:
$G_\text{exp}^{\ast(i)} = \arg\min \mathcal{L} ( y_\text{exp}^{(i)}, \texttt{softmax}(z_\text{exp}^{(i)}) ),\quad
G_\text{cls}^{\ast} = \arg\min \mathcal{L} ( y_\text{cls}, \texttt{softmax}(z_\text{cls}) )$.

\section{Experimental results}
\label{sec:experiments}



\subsubsection{Data pre-processing}
We follow the common pre-processing procedure of the LIDC dataset\cite{armatoLungImageDatabase2011} summarised in \cite{baltatzisPitfallsSampleSelection2021}. Scans with slice thickness larger than $2.5\,mm$ are discarded for being unsuitable for lung cancer screening according to clinical guidelines \cite{kazerooniACRSTRPractice2014}, and the remaining scans are resampled to the resolution of $1\,mm^3$ isotropic voxels. Only nodules annotated by at least three radiologists are retained. Annotations for both malignancy and nodule attributes of each nodule are aggregated by the median value among radiologists. Malignancy score is binarised by a threshold of $3$: nodules with median malignancy score larger than $3$ are considered malignant, smaller than $3$ are considered benign, while the rest are excluded\cite{baltatzisPitfallsSampleSelection2021}. 
For each annotation, 
only a 2D patch of size $32 \times 32\,px$ is extracted from the central axial slice. Although an image is extracted for each annotation, our training($70\%$)/testing($30\%$) split is on nodule level to ensure no image of the same nodule exists in both training and testing sets. This results in $276/242$ benign/malignant nodules for training and $108/104$ benign/malignant nodules for testing.

\subsubsection{Training settings}
Here we briefly state our training settings and refer to our code repository for further details.
The training of the feature extraction follows the suggestions in \cite{caronEmergingPropertiesSelfSupervised2021}. The encoders and prediction heads are trained for $300$ epochs with an AdamW optimiser 
and batch size $128$, starting from the weights pretrained unsupervisedly on ImageNet\cite{russakovskyImageNetLargeScale2015}. The learning rate is linearly scaled up to $0.00025$ during the first 10 epochs and then follows a cosine scheduler 
to decay till $10^{-6}$. The temperatures for the two branches are set to $\tau_\text{pri} = 0.04$, $\tau_\text{aux} = 0.1$. The momentum factor $\lambda$ is set to $0.9$, while $m$ is increased from $0.996$ to $1$ following a cosine scheduler.
The predictors $G_\text{exp}^{(i)}$ and $G_\text{cls}$ are jointly trained for $100$ epochs with SGD optimisers with momentum $0.9$ and batch size $128$. The learning rate follows a cosine scheduler with initial value $0.0005$ when using full annotation and $0.00025$ when using partial annotation.

The data augmentation for encoder training adapts from BYOL\cite{grillBootstrapYourOwn2020} and includes multi-crop as in \cite{caronUnsupervisedLearningVisual2020}. 
During the training of the predictors, the input images are 
augmented following previous works\cite{al-shabiLungNoduleClassification2019, baltatzisPitfallsSampleSelection2021} on the LIDC dataset. 

\subsection{Prediction performance of nodule attributes and malignancy}

\begin{table}[tbp]
    \centering
    \caption{\textbf{Prediction accuracy ($\%$) of nodule attributes and malignancy.}
    The best in each column is \textbf{bolded} for full/partial annotation respectively.
    Dashes (-) denote values not reported by the compared methods. 
    Results of our proposed cRedAnno are \colorbox[HTML]{E2EFD9}{highlighted}. 
    Observe that cRedAnno in almost all cases outperforms other methods in nodule attributes significantly, and also shows robustness w.r.t. configurations, meanwhile using the fewest nodules and no additional information.
    }
    \label{tab:res}
    \resizebox{\textwidth}{!}{%
    \begin{threeparttable}
        \begin{tabular}{lcccccccccc}
            \hline
             & \multicolumn{7}{c}{\textbf{Nodule attributes}} & \multicolumn{1}{l}{} & \multicolumn{1}{l}{} & \multicolumn{1}{l}{} \\ \cline{2-8}
             & \textbf{Sub} & \textbf{Cal} & \textbf{Sph} & \textbf{Mar} & \textbf{Lob} & \textbf{Spi} & \textbf{Tex} & \multicolumn{1}{l}{\multirow{-2}{*}{Malignancy}} & \multicolumn{1}{l}{\multirow{-2}{*}{\#nodules}} & \multicolumn{1}{l}{\multirow{-2}{*}{\begin{tabular}[c]{@{}c@{}}No additional \\ information\end{tabular}}} \\ \hline

            \multicolumn{11}{l}{Full annotation} \\
            \textbf{HSCNN}\cite{shenInterpretableDeepHierarchical2019} & 71.90 & 90.80 & 55.20 & 72.50 & - & - & 83.40 & 84.20 & 4252 & \xmark\tnote{c} \\
            \textbf{X-Caps}\cite{lalondeEncodingVisualAttributes2020} & 90.39 & - & 85.44 & 84.14 & 70.69 & 75.23 & 93.10 & 86.39 & 1149 & \cmark \\
            \textbf{MSN-JCN}\cite{chenEndtoEndMultiTaskLearning2021} & 70.77 & 94.07 & 68.63 & 78.88 & \textbf{94.75} & 93.75 & 89.00 & 87.07 & 2616 & \xmark\tnote{d} \\
            \textbf{MTMR}\cite{liuMultiTaskDeepModel2020} & - & - & - & - & - & - & - & \textbf{93.50} & 1422 & \xmark\tnote{e} \\
            \rowcolor[HTML]{E2EFD9} 
            \textbf{cRedAnno (50-NN)} & 94.93 & 92.72 & 95.58 & 93.76 & 91.29 & 92.72 & 94.67 & 87.52 & \cellcolor[HTML]{E2EFD9} & \cellcolor[HTML]{E2EFD9} \\
            \rowcolor[HTML]{E2EFD9} 
            \textbf{cRedAnno (250-NN)} & \textbf{96.36} & 92.59 & 96.23 & 94.15 & 90.90 & 92.33 & 92.72 & 88.95 & \cellcolor[HTML]{E2EFD9} & \cellcolor[HTML]{E2EFD9} \\
            \rowcolor[HTML]{E2EFD9} 
            \textbf{cRedAnno (trained)} & 95.84 & \textbf{95.97} & \textbf{97.40} & \textbf{96.49} & 94.15 & \textbf{94.41} & \textbf{97.01} & 88.30 & \multirow{-3}{*}{\cellcolor[HTML]{E2EFD9}\textbf{730}} & \multirow{-3}{*}{\cellcolor[HTML]{E2EFD9}\cmark} \\ \hline
            
            \multicolumn{11}{l}{Partial annotation} \\
            \textbf{WeakSup}\cite{joshiLungNoduleMalignancy2021} \textbf{(1:5\tnote{a} )} & 43.10 & 63.90 & 42.40 & 58.50 & 40.60 & 38.70 & 51.20 & 82.40 &  &  \\
            \textbf{WeakSup}\cite{joshiLungNoduleMalignancy2021} \textbf{(1:3\tnote{a} )} & 66.80 & 91.50 & 66.40 & 79.60 & 74.30 & 81.40 & 82.20 & \textbf{89.10} & \multirow{-2}{*}{2558} & \multirow{-2}{*}{\xmark\tnote{f}} \\
            \rowcolor[HTML]{E2EFD9} 
            \textbf{cRedAnno (10\%\tnote{b}, 50-NN)} & 94.93 & 92.07 & \textbf{96.75} & \textbf{94.28} & \textbf{92.59} & 91.16 & \textbf{94.15} & 87.13 & \cellcolor[HTML]{E2EFD9} & \cellcolor[HTML]{E2EFD9} \\
            \rowcolor[HTML]{E2EFD9} 
            \textbf{cRedAnno (10\%\tnote{b}, 150-NN)} & \textbf{95.32} & 89.47 & 97.01 & 93.89 & 91.81 & 90.51 & 92.85 & 88.17 & \cellcolor[HTML]{E2EFD9} & \cellcolor[HTML]{E2EFD9} \\
            \rowcolor[HTML]{E2EFD9} 
            \textbf{cRedAnno (1\%\tnote{b}, trained)} & 91.81 & \textbf{93.37} & 96.49 & 90.77 & 89.73 & \textbf{92.33} & 93.76 & 86.09 & \multirow{-3}{*}{\cellcolor[HTML]{E2EFD9}\textbf{730}} & \multirow{-3}{*}{\cellcolor[HTML]{E2EFD9}\cmark} \\ \hline
        \end{tabular}%
        \begin{tablenotes}
            \item[a] $1:N$ indicates that $\frac{1}{1+N}$ of training samples have annotations on nodule attributes. (All samples have malignancy annotations.)
            \item[b] The proportion of training samples that have annotations on nodule attributes and malignancy.
            \item[c] 3D volume data are used.
            \item[d] Segmentation masks and nodule diameter information are used. Two other traditional methods are used to assist training.
            \item[e] All 2D slices in 3D volumes are used.
            \item[f] Multi-scale 3D volume data are used.
        \end{tablenotes}
    \end{threeparttable}
    }
\end{table}

Two categories of experiments are conducted to evaluate the prediction accuracy of both malignancy and each nodule attribute: 
\begin{inlinelist}
    \item using k-NN classifiers to assign a label to each feature $f$ extracted from testing images by comparing the dot-product similarity with the ones extracted from training images, without any training;
    \item predicting via trained predictors $G_\text{exp}^{(i)}$ and $G_\text{cls}$. For simplicity, predictors $G_\text{exp}^{(i)}$ and $G_\text{cls}$ only use one linear layer.
\end{inlinelist}
Both k-NN classifier and trained predictors are evaluated with full/partial annotation, where partial annotation means only a certain percentage of training samples have annotations on nodule attributes and malignancy.
Each annotation is considered independently \cite{shenInterpretableDeepHierarchical2019}. The predictions of nodule attributes are considered correct if within $\pm1$ of aggregated radiologists' annotation \cite{lalondeEncodingVisualAttributes2020}. Attribute ``internal structure" is excluded from the results because its heavily imbalanced classes are not very informative \cite{shenInterpretableDeepHierarchical2019, lalondeEncodingVisualAttributes2020, chenEndtoEndMultiTaskLearning2021, liuMultiTaskDeepModel2020, joshiLungNoduleMalignancy2021}.

\begin{figure}[tbp]
    \centering
    \includegraphics[width=0.45\textwidth]{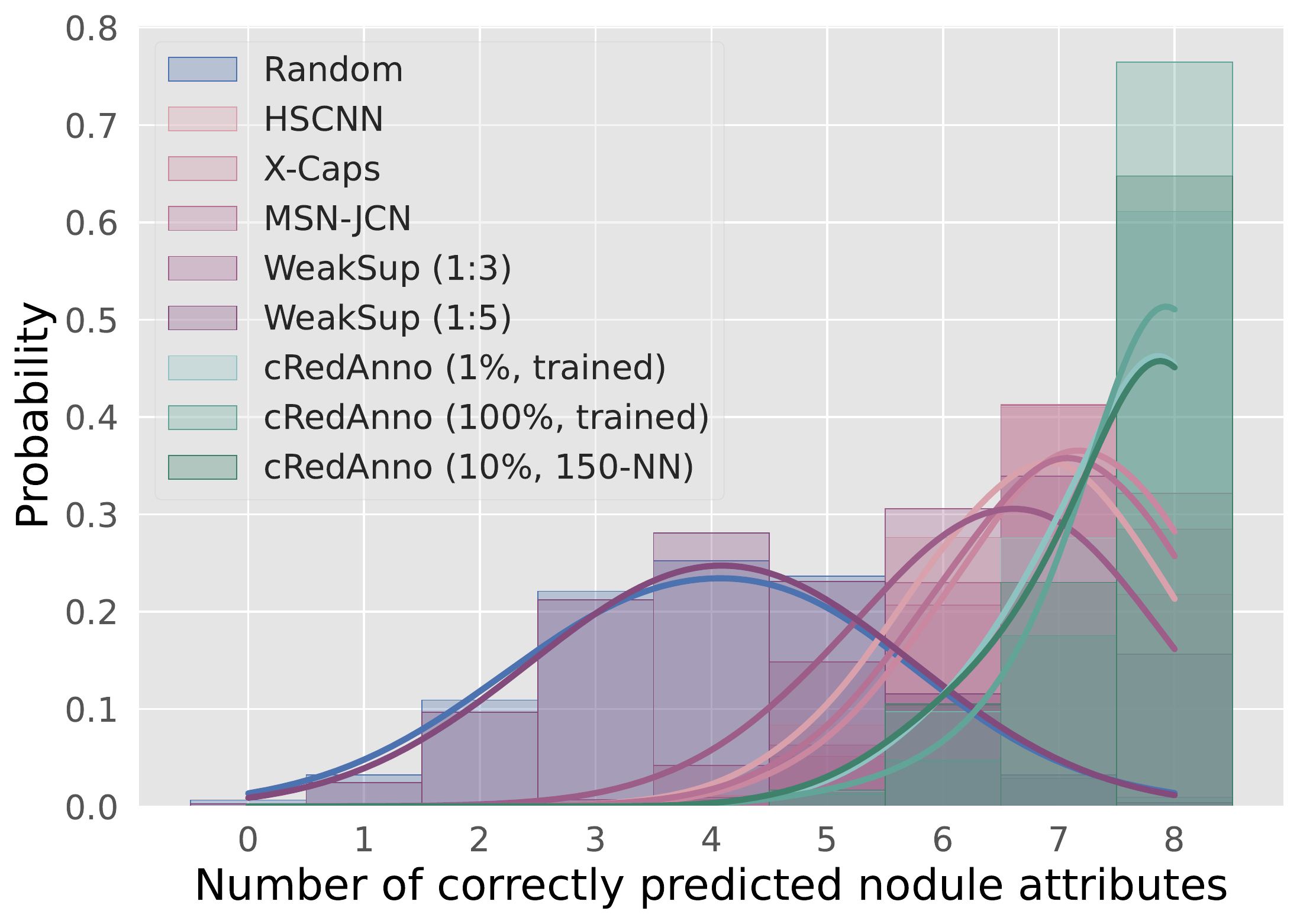}

    \caption{\textbf{Probability of the number of correctly predicted nodule attributes.}
    The probabilities of other methods are calculated using their reported prediction accuracy of individual nodule attributes, as in Tab.~\ref{tab:res}, where not-reported values are all assumed to be \textbf{100\% accuracy}.
    Observe that cRedAnno shows a significantly larger probability of simultaneously predicting all $8$ nodule attributes correctly.
    }
    \label{fig:res_prob_ftr}
\end{figure}

The overall prediction performance is summarised in Tab.~\ref{tab:res}, comparing with the state-of-the-art. 
In summary, the results show that our proposed approach can reach simultaneously high accuracy in predicting malignancy and all nodule attributes. This increases the trustworthiness of the model significantly and has not been achieved by previous works. 
More specifically, when using only $1\%$ annotated samples, our approach achieves comparable or much higher accuracy compared with all previous works in predicting the nodule attributes. Meanwhile, the accuracy of predicting malignancy approaches X-Caps \cite{lalondeEncodingVisualAttributes2020} and already exceeds HSCNN \cite{shenInterpretableDeepHierarchical2019}, which uses 3D volume data. Note that in this case we significantly outperform WeakSup(1:5) \cite{joshiLungNoduleMalignancy2021}, which uses $100\%$ malignancy annotations and $16.7\%$ nodule attribute annotations. 
When using full annotation, our approach outperforms most of the other compared explainable methods in predicting malignancy and all nodule attributes, except ``lobulation", where ours is merely worse by absolute $0.6\%$ accuracy. It is worth mentioning that even in this case, we still use the fewest samples: only $518$ among the $730$ nodules are used for training.
In addition, the consistent decent performance also indicates that our approach is reasonably robust w.r.t. to the value $k$ in k-NN classifiers.

To further validate the prediction performance of nodule attributes, for visual clarity, we select $3$ representative configurations of our proposed approach and compare them with others in Fig.~\ref{fig:res_prob_ftr}. 
It can be clearly seen that using our approach, approximately $90\%$ nodules have at least $7$ attributes correctly predicted. In contrast, WeakSup(1:5) although reaches over $82.4\%$ accuracy in malignancy prediction, shows no significant difference compared to random guesses in predicting nodule attributes -- this shows the opposite of trustworthiness.

\subsection{Analysis of extracted features in learned space}

\begin{figure}[tbp]
    \centering
    \subfloat[Malignancy]{
        \includegraphics[width=0.23\textwidth]{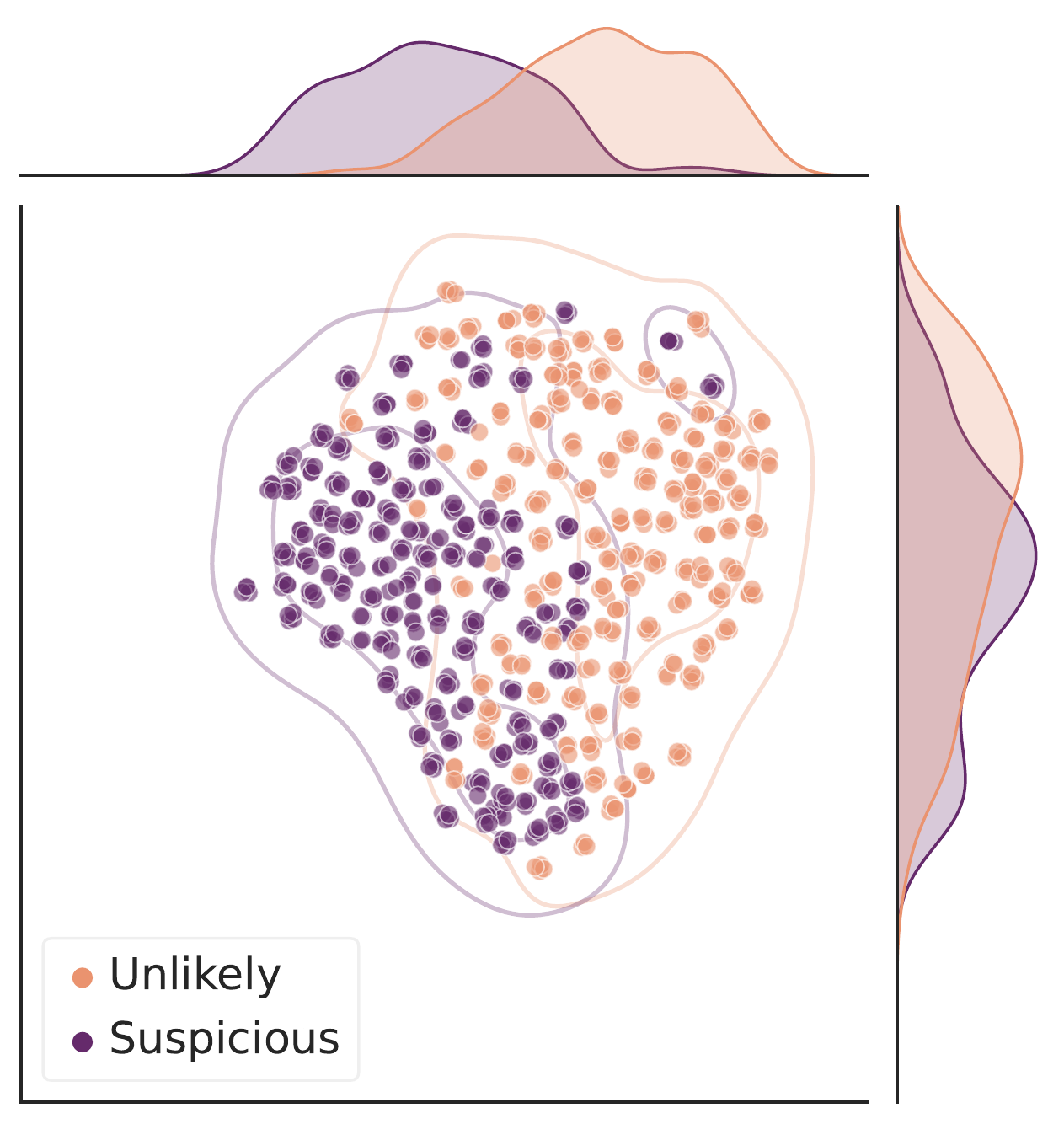}
        \label{fig:tsne_mal}
        }\hfil
    \subfloat[Subtlety]{
        \includegraphics[width=0.23\textwidth]{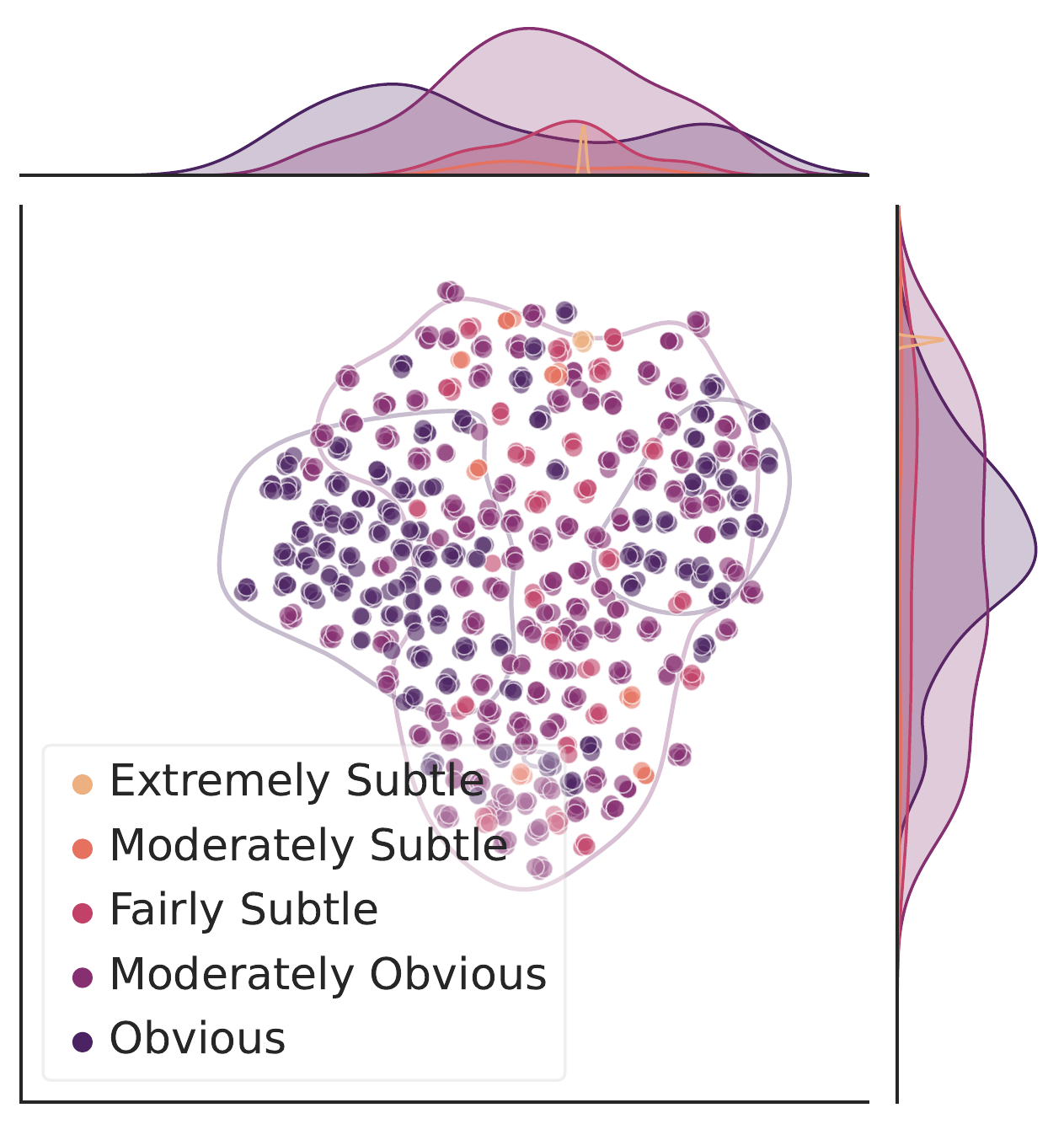}
        \label{fig:tsne_sub}
        }\hfil
    \subfloat[Calcification]{
        \includegraphics[width=0.23\textwidth]{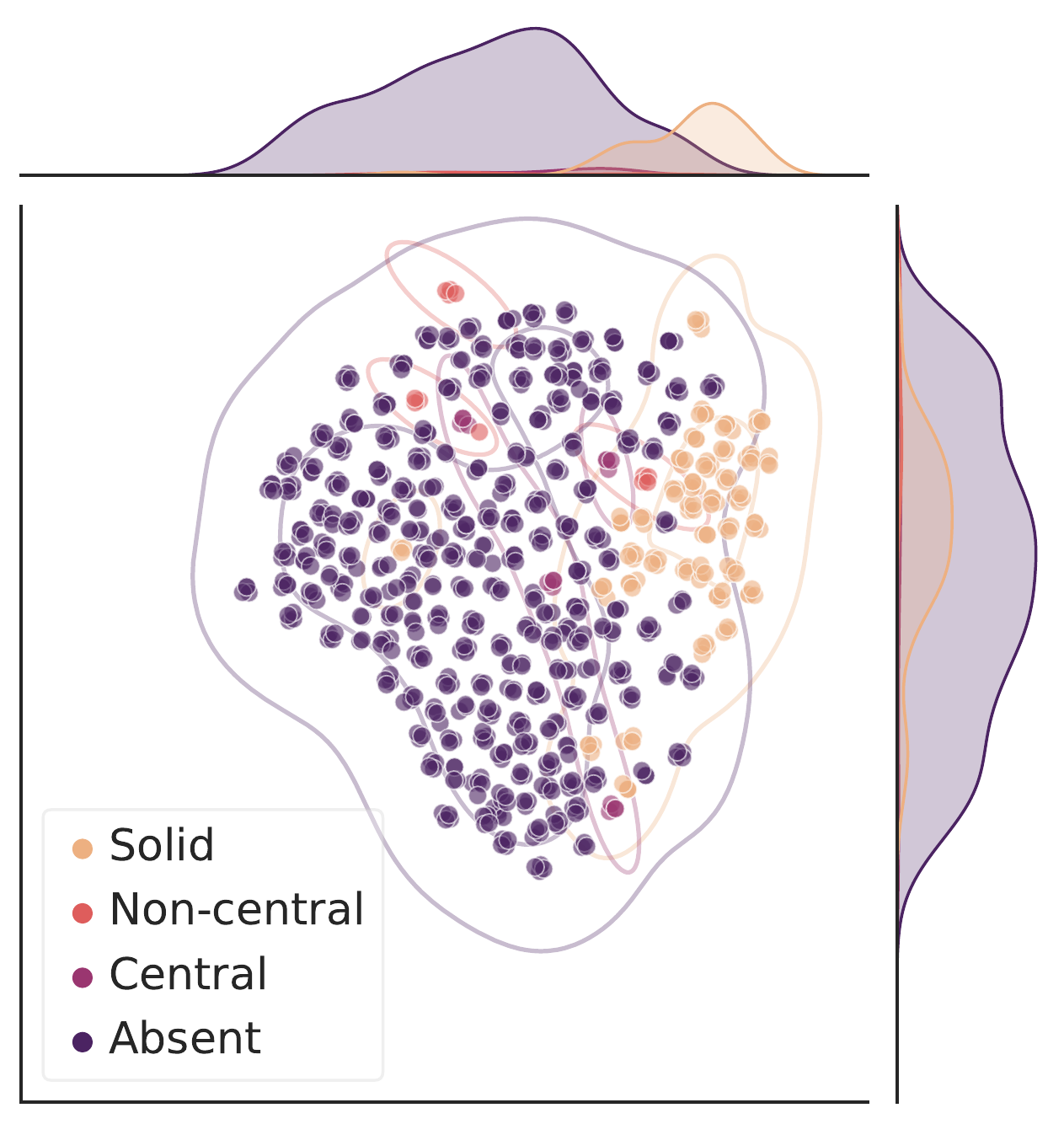}
        \label{fig:tsne_cal}
        }\hfil
    \subfloat[Sphericity]{
        \includegraphics[width=0.23\textwidth]{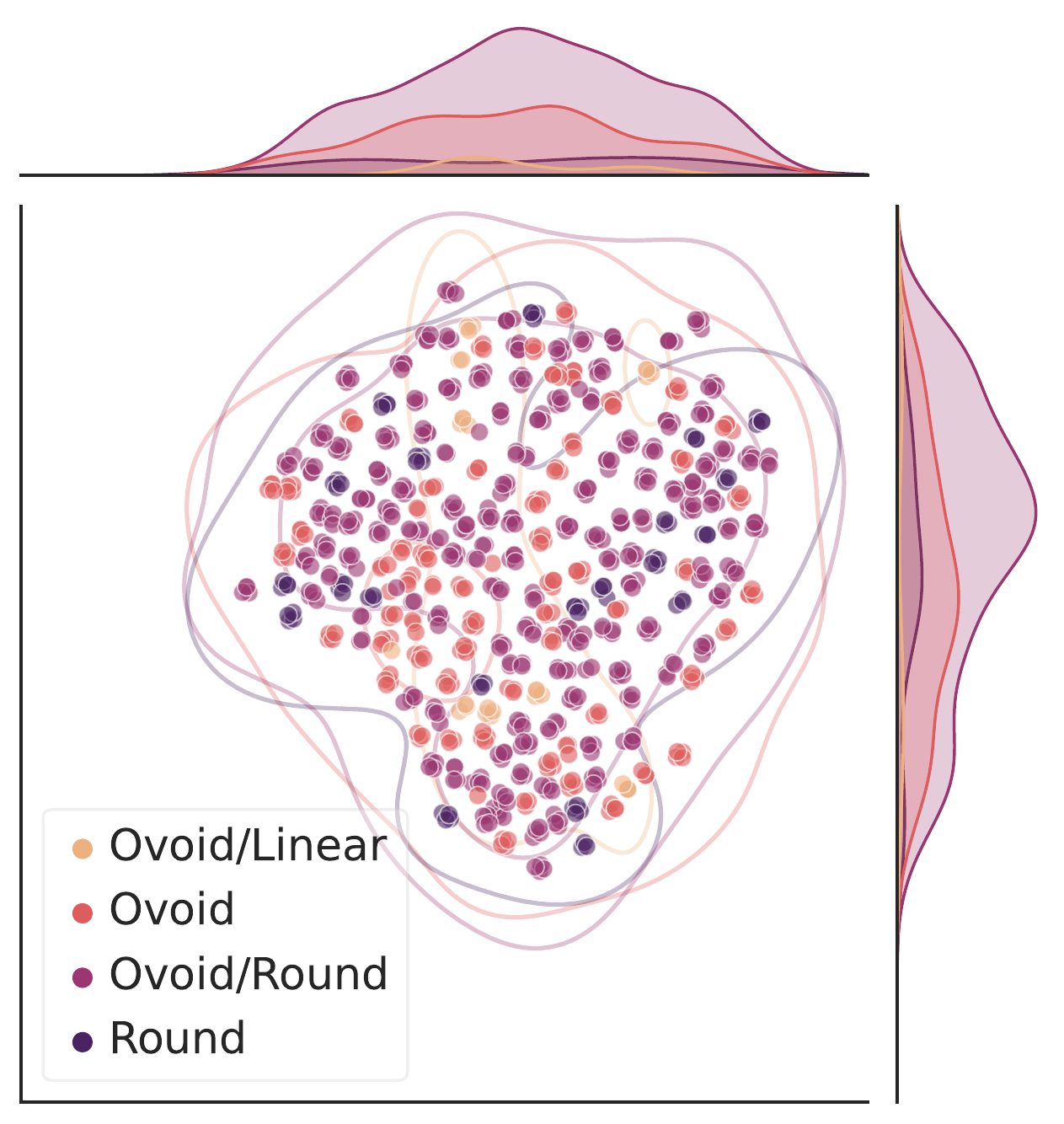}
        \label{fig:tsne_sph}
        }

    \subfloat[Margin]{
        \includegraphics[width=0.23\textwidth]{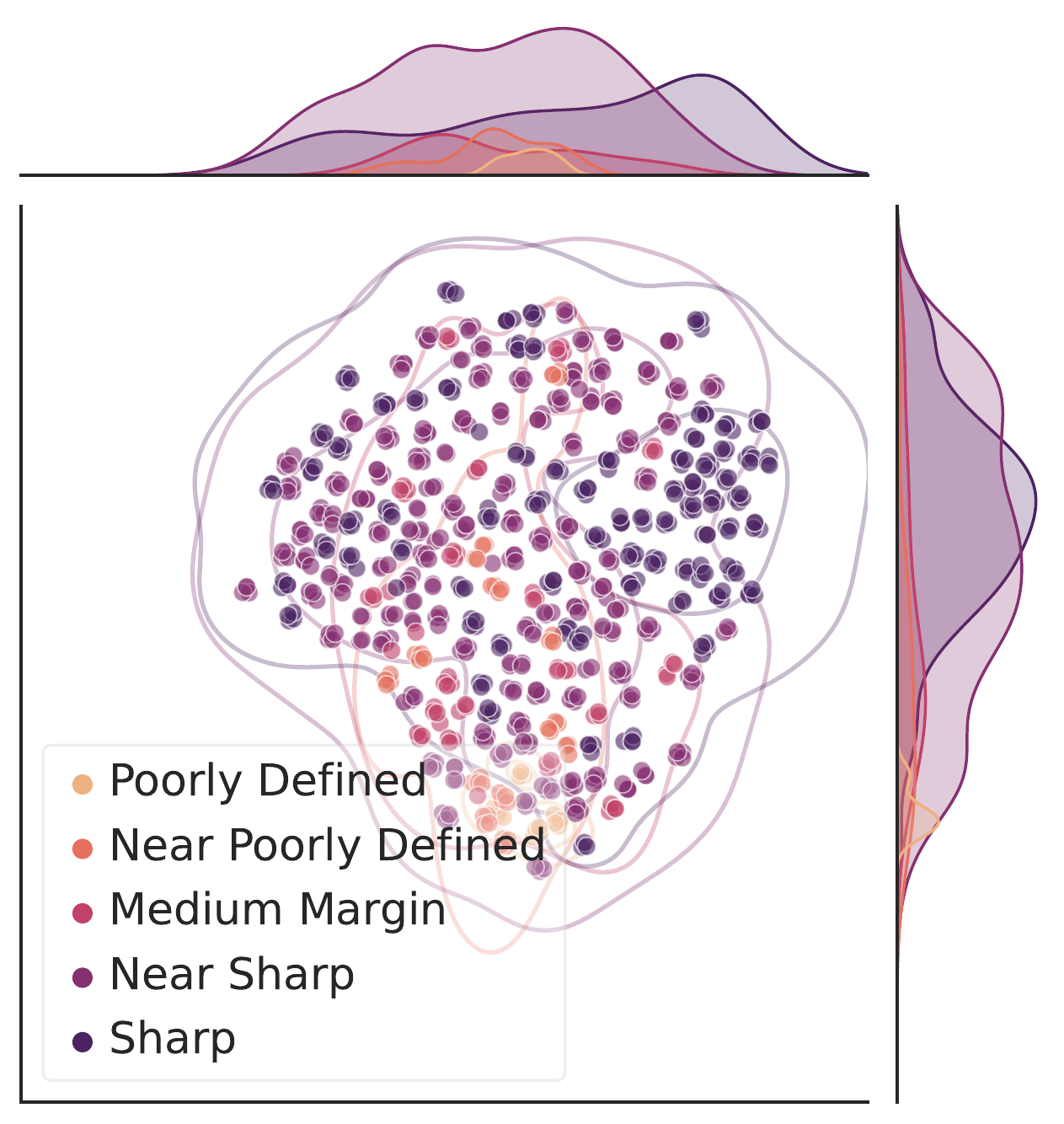}
        \label{fig:tsne_mar}
        }\hfil
    \subfloat[Lobulation]{
        \includegraphics[width=0.23\textwidth]{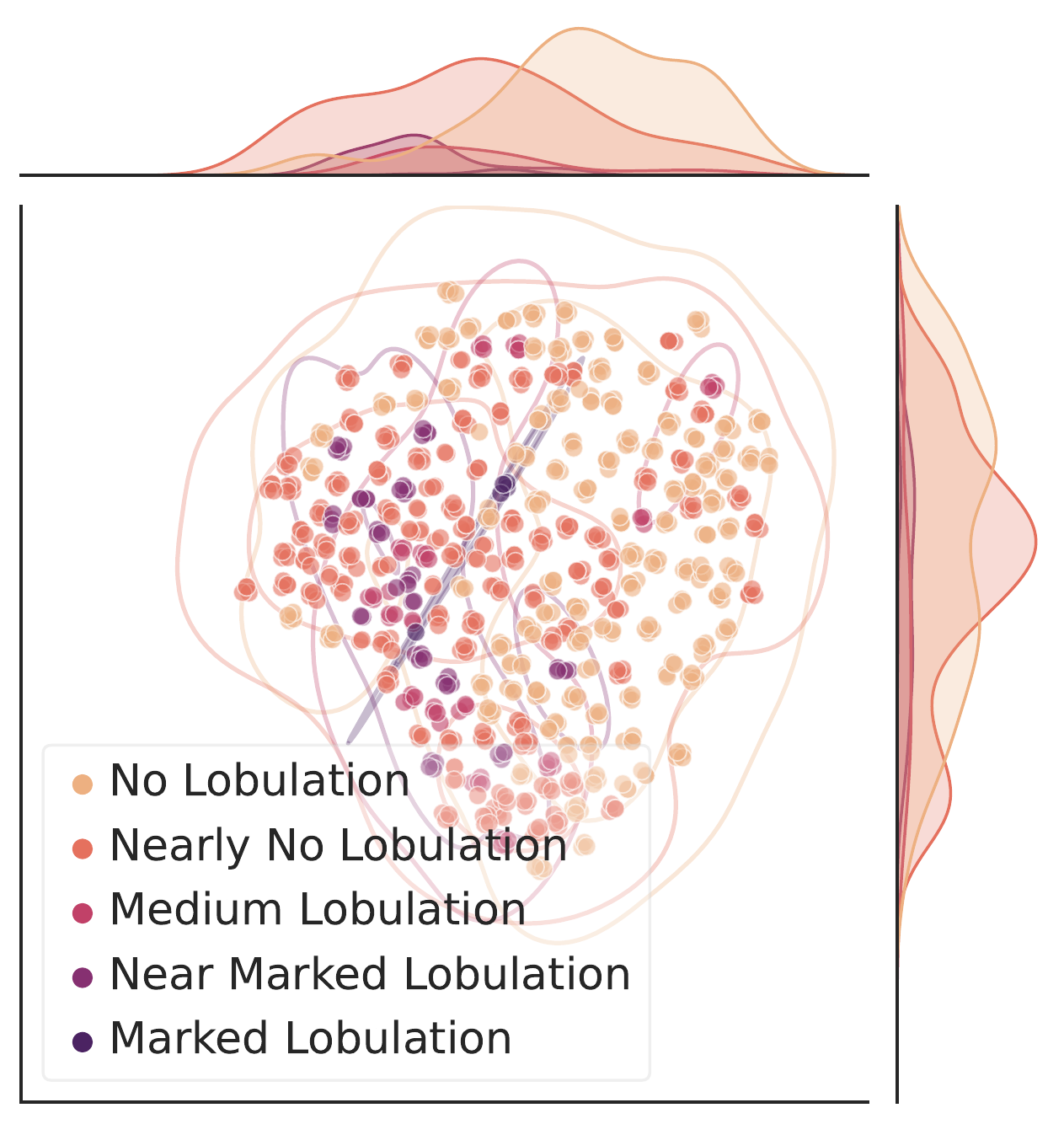}
        \label{fig:tsne_lob}
        }\hfil
    \subfloat[Spiculation]{
        \includegraphics[width=0.23\textwidth]{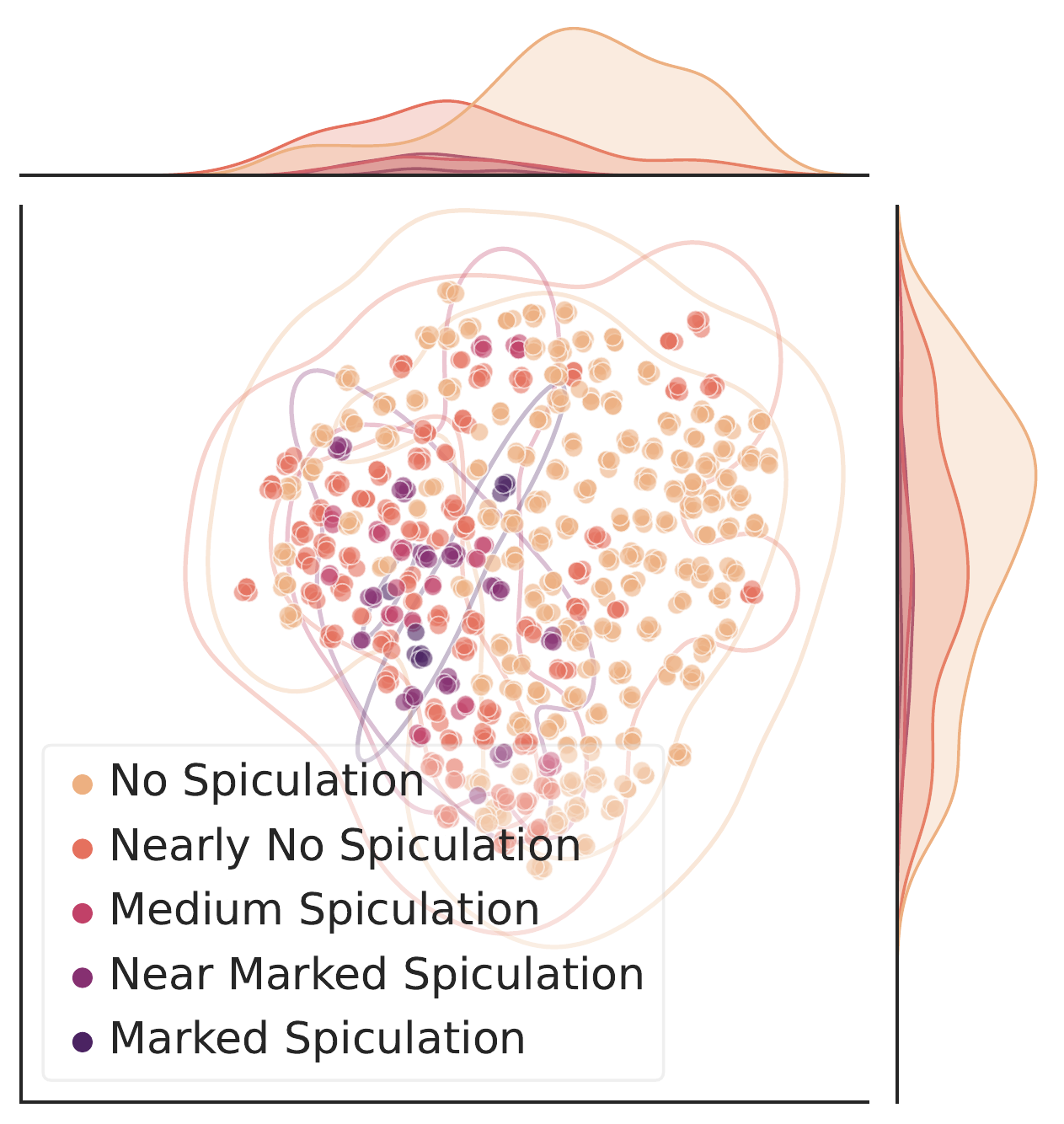}
        \label{fig:tsne_spi}
        }\hfil
    \subfloat[Texture]{
        \includegraphics[width=0.23\textwidth]{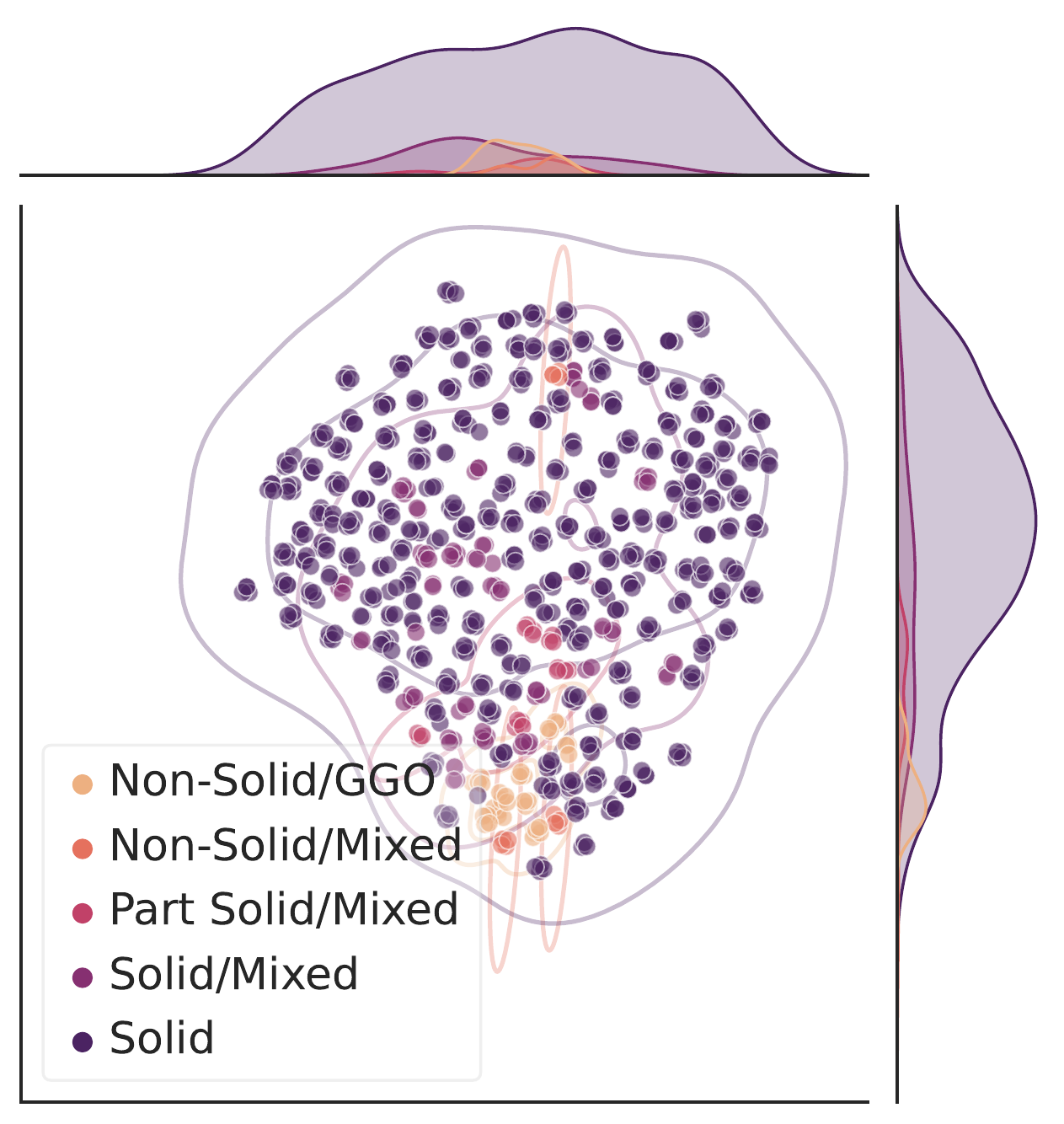}
        \label{fig:tsne_tex}
        }

    \caption{\textbf{t-SNE visualisation of features extracted from testing images.}
    Data points are coloured using ground truth annotations.
    Malignancy shows highly separable in the learned space, and correlates with the clustering in each nodule attribute.
    }
    \label{fig:res_tsne}
\end{figure}

We hypothesise the superior performance of our proposed approach can attribute to the extracted features. So we use t-SNE \cite{maatenVisualizingDataUsing2008} to further visualise the learned feature. Feature $f$ extracted from each testing image is mapped to a data point in 2D space. Fig.~\ref{fig:tsne_mal} to \ref{fig:tsne_tex} correspond to these data points coloured by the ground truth annotations of malignancy to nodule attribute ``texture", respectively. 
Fig.~\ref{fig:tsne_mal} shows that the samples are reasonably linear-separable between the benign/malignant samples even in this dimensionality-reduced 2D space. This provides evidence of our good performance.

Furthermore, the correlation between the nodule attributes and malignancy can be found intuitively in Fig.~\ref{fig:res_tsne}. 
For example, the cluster in Fig.~\ref{fig:tsne_cal} indicates that solid calcification contributes negatively to malignancy. Similarly, the clusters in Fig.~\ref{fig:tsne_mar} and Fig.~\ref{fig:tsne_tex} indicate that poorly defined margin correlates with non-solid texture, and both of these contribute positively to malignancy. These findings are in accord with the diagnosis process of radiologists\cite{vlahosLungCancerScreening2018} and thus further support the trustworthiness of the proposed approach.

\subsection{Ablation study}

\begin{figure}[tbp]
    \CenterFloatBoxes
    \begin{floatrow}
    \ttabbox[\FBwidth]{%
    \caption{\textbf{Accuracy of malignancy prediction ($\%$).} All annotations are used during training. The highest accuracy is \textbf{bolded}. The result of our proposed setting is \textcolor[HTML]{2F847C}{highlighted}.
    Only cRedAnno and conventional end-to-end trained CNN achieve higher than $85\%$ accuracy. 
    }
    \label{tab:res_ablation}
    \resizebox{0.95\linewidth}{!}{%
    \begin{threeparttable}
        \begin{tabular}{ccccc}
        \hline
        \textbf{Arch} &
          \textbf{\#params} &
          \textbf{\begin{tabular}[c]{@{}c@{}}Training\\ strategy\end{tabular}} &
          \textbf{\begin{tabular}[c]{@{}c@{}}ImageNet\\ pretrain\end{tabular}} &
          \textbf{Acc} \\ \hline
        \multirow{3}{*}{ResNet-50} & \multirow{3}{*}{23.5M} & \textcolor[HTML]{B84878}{end-to-end} & \textcolor[HTML]{B84878}{\xmark} & \textcolor[HTML]{B84878}{86.74\tnote{$\ast$}} \\
                                   &                        & two-stage  & \xmark & 70.48 \\
                                   &                        & two-stage  & \cmark & 70.48 \\
        \multirow{3}{*}{ViT}       & \multirow{3}{*}{21.7M} & end-to-end & \xmark & 64.24 \\
                                   &                        & two-stage  & \xmark & 79.19 \\
                                       &                        & \textcolor[HTML]{2F847C}{two-stage}  & \textcolor[HTML]{2F847C}{\cmark} & \textcolor[HTML]{2F847C}{\textbf{88.30}} \\ \hline
        \end{tabular}%
        \begin{tablenotes}
            \item[$\ast$] This is a representative setting and performance of previous works using CNN architecture.
        \end{tablenotes}
    \end{threeparttable}
    }
    }
    \hfil
    \killfloatstyle
    \ffigbox[\FBwidth]{%
      \includegraphics[width=0.97\linewidth]{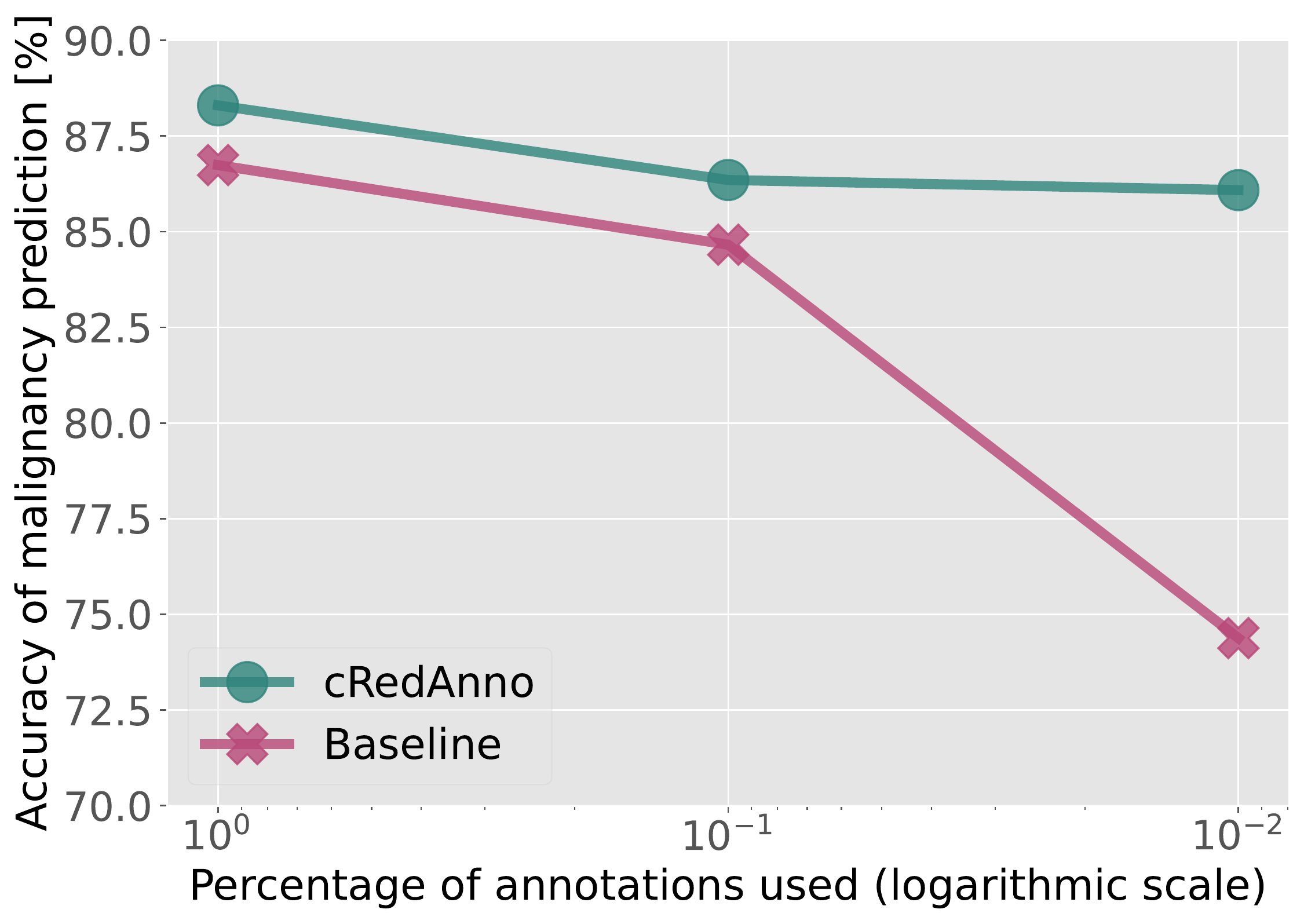}%
    }{%
      \caption{\textbf{Annotation reduction.} 
      Line colours correspond to settings in Tab.~\ref{tab:res_ablation}: ``\textcolor[HTML]{B84878}{Baseline}" uses ResNet-50 architecture and is trained end-to-end from random initialisation.
      cRedAnno shows strong robustness when annotation reduced.
      }%
      \label{fig:res_annoreduce}
    }
    \end{floatrow}
\end{figure}



\subsubsection{Validation of components}
We ablate our proposed approach by comparing with different architectures for encoders $E$, training strategies, and whether to use ImageNet-pretrained weights. The results in Tab.~\ref{tab:res_ablation} show that ViT architecture benefits more from the self-supervised contrastive training compared to ResNet-50 as a CNN representative. This observation is in accord with the findings in \cite{chenEmpiricalStudyTraining2021, caronEmergingPropertiesSelfSupervised2021}. ViT's lowest accuracy in end-to-end training reiterates its requirement for a large amount of training data \cite{dosovitskiyImageWorth16x162020}. Starting from the ImageNet-pretrained weights is also shown to be helpful for ViT but not ResNet-50, probably due to ViT's lack of inductive bias needs far more than hundreds of training samples to compensate \cite{dosovitskiyImageWorth16x162020}, especially for medical images.
In summary, only the proposed approach and conventional end-to-end training of ResNet-50 achieve higher than $85\%$ accuracy of malignancy prediction. 

\subsubsection{Annotation reduction}
We further plot the malignancy prediction accuracy of the aforementioned winners as the annotations are reduced on a logarithmic scale.
As shown in Fig.~\ref{fig:res_annoreduce}, cRedAnno demonstrates strong robustness w.r.t. annotation reduction. The accuracy of the end-to-end trained ResNet-50 model decreases rapidly to $74.38\%$ when annotations reach only $1\%$. In contrast, the proposed approach still remains at $86.09\%$ accuracy, meanwhile high accuracy for predicting nodule attributes, as shown in Tab.~\ref{tab:res}.

\section{Conclusion}
\label{sec:conclustion}
In this study, we propose cRedAnno to considerably reduce the annotation need in predicting malignancy, meanwhile explaining nodule attributes for lung nodule diagnosis. 
Our experiments show that even with only $1\%$ annotation, cRedAnno can reach similar or better performance in predicting malignancy compared with state-of-the-art methods using full annotation, and significantly outperforms them in predicting nodule attributes. 
In addition, our proposed approach is the first to reach over $94\%$ accuracy in predicting all nodule attributes simultaneously.
Visualisation of our extracted features provides novel evidence that in the learned space, the clustering of nodule attributes and malignancy is in accord with clinical knowledge of lung nodule diagnosis.
Yet the limitations of this approach remain in its generalisability to be validated in other medical image analysis problems.

%
%

\bibliographystyle{splncs04}
\bibliography{Explainability}

\end{document}